\documentclass[runningheads]{llncs}
\usepackage[T1]{fontenc}
\usepackage{graphicx}
\usepackage{breqn}
\usepackage{multirow}
\begin{document}
\title{Weakly Semi-Supervised Detection in Lung Ultrasound Videos}
%
%
%

\author{
Jiahong Ouyang\thanks{Work completed during internship at Philips Research North America.} \inst{1}  \and
Li Chen\thanks{Corresponding author. Email: li.chen_1@philips.com} \inst{2} \and
Gary Y. Li\inst{2} \and
Naveen Balaraju\inst{2} \and
Shubham Patil \inst{2} \and
Courosh Mehanian \inst{3} \and
Sourabh Kulhare \inst{3} \and
Rachel Millin \inst{3} \and
Kenton W. Gregory \inst{4} \and
Cynthia R. Gregory \inst{4} \and
Meihua Zhu \inst{4} \and
David O. Kessler \inst{5} \and
Laurie Malia \inst{5} \and
Almaz Dessie \inst{5} \and
Joni Rabiner \inst{5} \and
Di Coneybeare \inst{5} \and
Bo Shopsin \inst{6} \and
Andrew Hersh \inst{7} \and
Cristian Madar \inst{8} \and
Jeffrey Shupp \inst{9} \and
Laura S. Johnson \inst{9} \and
Jacob Avila \inst{10} \and
Kristin Dwyer \inst{11} \and
Peter Weimersheimer \inst{12} \and
Balasundar Raju  \inst{2} \and
Jochen Kruecker \inst{2} \and
Alvin Chen \inst{2}}
\authorrunning{J. Ouyang et al.}
\institute{Stanford University, Stanford CA, USA \and Philips Research North America, Cambridge MA, USA \and 
Global Health Laboratories, Bellevue, WA, USA \and
Oregon Health \& Science University, Portland, OR, USA \and
Columbia University Medical Center, New York, NY, USA \and
New York University, New York, NY, USA \and
Brooke Army Medical Center, San Antonio, TX, USA \and
Tripler Army Medical Center, Honolulu, HI, USA \and
MedStar Washington Hospital Center, Washington, DC, USA \and
University of Kentucky, Lexington, KY, USA \and
Warren Alpert Medical School of Brown University, Providence, RI, USA \and
University of Vermont Larner College of Medicine, Burlington, VT, USA \\}

\maketitle              
\begin{abstract}
Frame-by-frame annotation of bounding boxes by clinical experts is often required to train fully supervised object detection models on medical video data. We propose a method for improving object detection in medical videos through weak supervision from video-level labels. More concretely, we aggregate individual detection predictions into video-level predictions and extend a teacher-student training strategy to provide additional supervision via a video-level loss. We also introduce improvements to the underlying teacher-student framework, including methods to improve the quality of pseudo-labels based on weak supervision and adaptive schemes to optimize knowledge transfer between the student and teacher networks. We apply this approach to the clinically important task of detecting lung consolidations (seen in respiratory infections such as COVID-19 pneumonia) in medical ultrasound videos. Experiments reveal that our framework improves detection accuracy and robustness compared to baseline semi-supervised models, and improves efficiency in data and annotation usage.

\end{abstract}


\section{Introduction}

Despite the remarkable performance of deep learning networks for object detection and other computer vision tasks \cite{dlus1,dlus2,dlus3}, most models rely on large-scale annotated training examples, which are often unavailable or burdensome to generate in the medical imaging domain. This is especially true for video-based imaging modalities such as medical ultrasound, where frame-by-frame annotation of bounding boxes or other localization labels is extremely time-consuming and costly, and even more so if annotations must be done by clinical experts. 


Reducing the annotation burden of training object detectors on medical images has been a focus of much recent work. Semi-supervised and weakly supervised approaches have been proposed to address the annotation challenge, where unlabeled or inexactly/inaccurately labeled data are used to supplement training, often in combination with a small amount of fully labeled data \cite{Peng2021,Jiao2022,liu2021unbiased,shao2022deep,zhang2021weakly,bakalo2021weakly}. Examples of weak supervision for object detection include point annotations \cite{zhang2022group,ge2022point,wang2022omni,ren2020ufo,chai2022orf} and image-level class labels \cite{wang2022omni,ren2020ufo,meethal2022semi}, both of which are applied on individual image frames. However, even these methods of weak supervision may not be practical in the video domain, where hundreds or thousands of image frames requiring interpretation may be collected in a single clinical exam.

In this work, we propose a weakly semi-supervised framework for training object detection models based on video-level supervision, where only a single label is provided for each video. Video-level labels represent a significantly weaker form of supervision than instance- or image-level labels but can be generated much more efficiently. Our approach extends teacher-student models adopted for semi-supervised object detection \cite{liu2021unbiased} to the weakly semi-supervised video-based detection task. Our main contributions are as follows:

\begin{enumerate}
\item We introduce a simple mechanism during teacher-student training which aggregates individual detections from the teacher (pseudo-labels) into video-level confidence predictions. This allows video-level weak supervision using any standard classification loss.
\item We improve the reliability of pseudo-labels generated during the mutual learning stage by introducing techniques to re-weigh pseudo-labels based on video-level weak supervision.
\item We investigate the learning dynamics between the teacher and student, and propose several improvements to the underlying teacher-student mechanism to increase training stability. These include a method to better initialize models in the "burn-in" stage and a set of adaptive updating schemes to optimize knowledge transfer bidirectionally during mutual learning.
\end{enumerate}

We demonstrate the effectiveness of our approach on the task of detecting lung consolidations in medical ultrasound videos, which is an important step in aiding diagnosis and management of patients with respiratory infections such as bacterial and viral pneumonia (including COVID-19 infection). Computer-aided detection of lung consolidation in medical ultrasound is a uniquely challenging problem where the appearance of pathology varies dramatically across disease types, patient populations, and training levels of the personnel acquiring images. Experimental results on a large, multi-center, clinical ultrasound dataset for lung consolidation demonstrate that the proposed framework leads to improved detection performance, robustness, and training stability compared to existing semi-supervised methods.

\section{Related Work}


\noindent\textbf{Semi-Supervised Object Detection:}
Semi-supervised object detection aims to utilize large amounts of unlabeled data together with a small set of labeled data. These efforts generally fall into two categories: (1) consistency regularization, which regularizes the prediction of the detector for images undergoing different augmentations \cite{jeong2019consistency,jeong2021interpolation}, and (2) pseudo-labeling, where a teacher model is trained on labeled data to generate pseudo-labels for unlabeled data, and a student model is then trained on both the labeled and pseudo-labeled data \cite{liu2021unbiased,zhou2021instant,xu2021end,tang2021humble,wang2021data}. Unbiased Teacher (UBT) \cite{liu2021unbiased} is one of the state-of-the-art methods in this category. Our work is inspired by the framework of UBT, but extends the method to the weakly semi-supervised scenario, where we leverage frame-level pseudo-labels and video-level weak supervision. 


\noindent\textbf{Weakly Semi-supervised Object Detection:}
Weakly semi-supervised object detection is usually based on instance-level weak supervision, e.g., a point on the object \cite{zhang2022group,ge2022point,wang2022omni,ren2020ufo,chai2022orf,ji2022point}, or image-level supervision, e.g., the class of the image \cite{wang2022omni,ren2020ufo,meethal2022semi}. Video-level labels are a significantly weaker supervisory signal than instance- or frame-level labels, since the only information provided is that the object class exists somewhere in at least one frame of the video, but in which specific frame(s) and at what location(s) is unknown.

\section{Method}
The basic intuition behind our approach is to simultaneously learn from both frame-level and video-level labels to improve object detection performance. We denote a fully labeled set $\mathcal{D}_f$ and a weakly labeled set $\mathcal{D}_w$ as our training data. The fully labeled set $\mathcal{D}_f = \{x^f_i, y^f_i\}_{i=1}^{N_f}$ comprises a set of $N_f$ frames $x^f_i$ and their paired frame-level annotations $y^f_i$ (i.e., the coordinates of all bounding boxes present in each frame). The weakly labeled set $\mathcal{D}_w = \{x^w_{j, 1:T_j}, z^w_j\}_{j=1}^{N_w}$ consists of $N_w$ videos $x^w_{j, 1:T_j}$ and their video-level class labels $z^w_j$ (indicating whether a video contains at least one instance of the object class, which is annotated on the whole video of $T_j$ frames). Due to its high accuracy, computational efficiency, and capacity for real-time inference, YOLO-v5 \cite{yolov5} is adopted as the backbone detector. Non-maximum suppression (NMS) is applied on the outputs of YOLO-v5 as the final output of the detector to remove duplicate prediction. Here, we use $c$ to denote the confidence vector (a part of YOLO-v5's output) of all predicted boxes from a frame $x$.


\subsection{Teacher-Student Training}
\begin{figure}[t]
\centering
\includegraphics[width=0.9\textwidth]{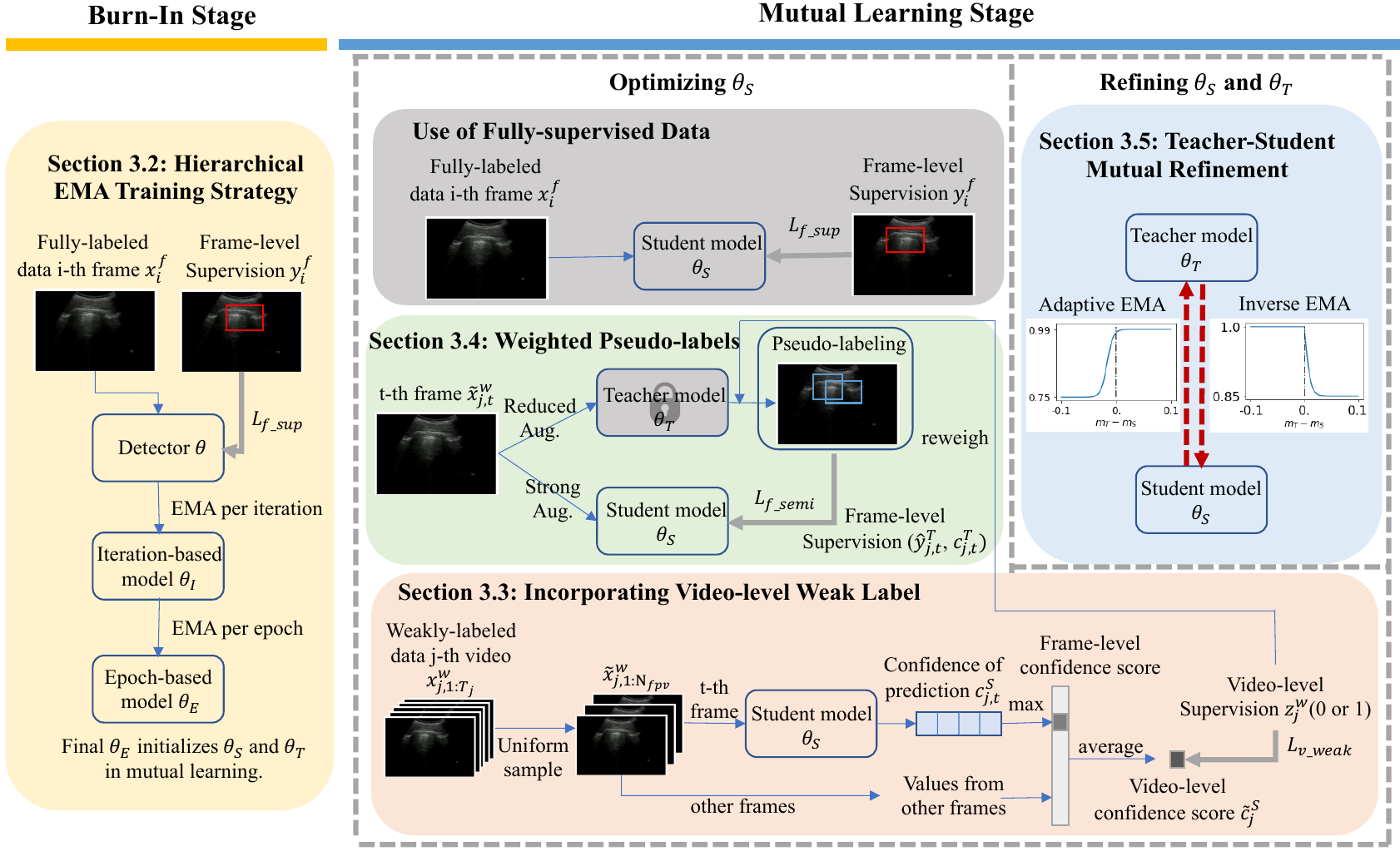}
\vspace{-15pt}
\caption{Overview of the proposed method.} 
\label{fig:overview}
\vspace{-5pt}
\end{figure}

We adopt the UBT framework from \cite{liu2021unbiased} for the weakly semi-supervised detection task. UBT uses two training stages: a burn-in stage for model initialization and a mutual learning stage for teacher-student training (Fig. \ref{fig:overview}). 

\noindent\textbf{Burn-In Stage:} The burn-in stage aims to initialize the detector in a supervised manner using the fully labeled data $\mathcal{D}_f$.
The original burn-in method in UBT adopts model weights after a fixed number of early training epochs. However, we observed that detection performance can vary dramatically during training. Here, we improve the training stability during burn-in by applying hierarchical exponential moving average (EMA) updates at each iteration and epoch (yellow block in Fig. \ref{fig:overview}). Specifically, the iteration-based model $\theta_I$ is introduced to transfer knowledge from an initial detection model $\theta$ after each iteration (that is, the weights of $\theta_I$ are updated per batch). The epoch-based model $\theta_E$ is added to transfer information from $\theta_I$ after each epoch (weights updated after all batches). This is given by:
\vspace{-5pt}
\begin{equation}
\begin{cases}
    \theta_I \leftarrow \alpha_i \theta_I + (1-\alpha_i) \theta, & \textrm{for each iteration}\\
    \theta_E \leftarrow \alpha_e \theta_E + (1-\alpha_e) \theta_I, & \textrm{for each epoch}
\end{cases}
\vspace{-5pt}
\end{equation}
where $\alpha_i$ and $\alpha_e$ are the iteration- and epoch-based EMA keep rates, respectively. The EMA keep rates define a trade-off between the rate of knowledge being transferred from the preceding model versus the stability of the succeeding model. Major oscillations in model performance are seen while training $\theta_I$ alone, even with a carefully selected $\alpha_i$ \cite{yolov5}. In contrast, the addition of $\theta_E$ serves to stabilize the training and results in a better-initialized detector after burn-in.

\noindent\textbf{Mutual Learning Stage:} The mutual learning stage combines the fully labeled data $\mathcal{D}_f$ and weakly labeled data $\mathcal{D}_w$ for teacher-student training. Both the student and teacher models are initialized from the last checkpoint of $\theta_E$ trained in the burn-in stage. During mutual learning, the student $\theta_S$ is optimized via backpropagation using a combination of full, weak, and semi-supervised losses (see Sections 3.3 and 3.4), and the teacher $\theta_T$ is updated via a gradual EMA transfer of weights from the student. Analogous to the model updates in the burn-in stage, the student is updated with an iteration-based EMA during mutual learning, while the teacher is updated with an epoch-based EMA. An additional mutual refinement scheme is introduced to adaptively adjust the EMA keep rate of the teacher, as well as to conditionally allow transfer of weights back to the student (see Section 3.5). As mutual learning progresses, the accuracy and stability of pseudo-labels produced by the teacher are continuously improved, which in turn improves knowledge distillation to and from the student. At the end of the training, only the teacher $\theta_T$ is kept for evaluation and deployment.


\subsection{Weakly Semi-Supervised Learning}

\noindent\textbf{Frame-Level Full Supervision:}
For both burn-in and mutual learning, fully labeled data $\mathcal{D}_f$ are used in training, with supervision by a detection loss $\mathcal{L}_{f\_sup}$:
\vspace{-5pt}
\begin{equation}
\mathcal{L}_{f\_sup} = \sum_{i=1}^{N_f} \lambda_{coord} \mathcal{L}_{coord}(\mathcal{T}_s(x^f_i), y^f_i) + \lambda_{conf}  \mathcal{L}_{conf}(\mathcal{T}_s(x^f_i), y^f_i),
\vspace{-5pt}
\end{equation}
where $\mathcal{L}_{coord}$ is the bounding box coordinate error, and $\mathcal{L}_{conf}$ is the binary cross-entropy loss between predicted box confidences and corresponding box labels. $\lambda_{coord}$ and $\lambda_{conf}$ balance the two losses. $\mathcal{T}_s$ denotes the data augmentation.

\vspace{\baselineskip}
\noindent\textbf{Frame-Level Semi-Supervision:} 
In the mutual learning stage, the weakly labeled data $\mathcal{D}_w$ are added to allow frame-level semi-supervision based on pseudo-labels from the teacher $\theta_T$. Specifically, we generate sub-clips $\tilde{x}^w_{j, 1:N_{fpv}}$ by uniformly sampling $N_{fpv}$ frames from each video $x^w_{j, 1:T_j}$ in the weak dataset. These sub-clips are fed through the teacher with reduced augmentation $\mathcal{T}_r$ to obtain box predictions $\hat{y}^T_{j,1:N_{fpv}}$ with confidence scores $c^T_{j,1:N_{fpv}}$ from each frame.
Only predicted boxes with confidence above a threshold $\beta$ are kept as pseudo-labels. We use a second detection loss, similar to Eq. 2, to train the student $\theta_S$ against the teacher's pseudo-labels:
\vspace{-5pt}
\begin{equation}
\mathcal{L}_{f\_semi} = \sum_{j=1}^{N_w} \sum_{t=1}^{N_{fpv}} \left[ \lambda_{coord}  \mathcal{L}_{coord}(\mathcal{T}_s(x^w_{j,t}), \hat{y}^T_{j,t}) + \lambda_{conf}  \mathcal{L}_{conf}(\mathcal{T}_s(x^w_{j,t}), \hat{y}^T_{j,t}) \right]
\vspace{-5pt}
\end{equation}

\noindent\textbf{Video-Level Weak Supervision:} 
Finally, we utilize the confidence of boxes predicted by the student $c^S_{j, 1:N_{fpv}}$ to obtain a frame-level confidence by computing the maximum confidence among all detected boxes in the frame, i.e., $max(c^S_{j, t})$. A final video-level prediction $\tilde{c}^S_j$ is computed as the averaged frame-level confidence over the sub-clip, i.e., $\tilde{c}^S_j=\frac{1}{N_{fpv}} \sum_{t=1}^{N_{fpv}} max(c^S_{j,t})$.


Here, we apply a video-level binary cross-entropy classification loss to supervise the student $\theta_S$ against the video-level labels $z^w_j$ from the weak data $\mathcal{D}_w$:
\vspace{-5pt}
\begin{equation}
\mathcal{L}_{v\_weak} = - \sum_{j=1}^{N_w} \left[ z^w_j log(\tilde{c}^S_j) + (1-z^w_j)log(1-\tilde{c}^S_j) \right]
\vspace{-5pt}
\end{equation}

\noindent\textbf{Combined Loss:} 
The final loss function for training the student model combines the fully supervised detection loss $\mathcal{L}_{f\_sup}$, the frame-level semi-supervised detection loss $\mathcal{L}_{f\_semi}$, and the video-level weakly supervised loss $\mathcal{L}_{v\_weak}$:
\vspace{-5pt}
\begin{equation}
\mathcal{L}=\lambda_{f\_sup} \mathcal{L}_{f\_sup} + \lambda_{f\_semi} \mathcal{L}_{f\_semi} + \lambda_{v\_weak} \mathcal{L}_{v\_weak} 
\vspace{-5pt}
\end{equation}
with $\lambda_{f\_sup}$, $\lambda_{f\_semi}$, and $\lambda_{v\_weak}$ to balance the three loss components.

\subsection{Weighted Pseudo-labels}
A critical factor in the effectiveness of mutual learning is the quantity and quality of pseudo-labels from the teacher $\theta_T$. We propose two pseudo-label re-weighting techniques to increase the number of high-quality pseudo-labels during training. 

\noindent\textbf{Weakly Supervised Pseudo-label Filtering:} The first approach utilizes the weak video-level label $z^w_j$ to filter false pseudo-labels. For a negative video ($z^w_j=0$), we can simply remove all pseudo-labels from every frame, which could be considered as re-weighting the pseudo-label as 0. For a positive video ($z^w_j=1$), if no pseudo-label confidence exceeds $\beta$, we keep the pseudo-label with the highest confidence if it exceeds a lower threshold $\beta_l$, where $\beta_l<\beta$, which could be considered as re-weighting the pseudo-label from 0 to its confidence.


\noindent\textbf{Soft Pseudo-labels:} 
The second approach assigns a weight to each pseudo-label based on its prediction confidence. That is, we re-weigh the loss component for each pseudo-label $\hat{y}^T_{j,t}$ by the square of its confidence $(c^T_{j,t})^2$ to create "soft pseudo-labels" ($\hat{y}^T_{j,t}$, $c^T_{j,t}$). The semi-supervised detection loss is reformulated as:
\vspace{-5pt}
\begin{equation}
\begin{aligned}
\mathcal{L}_{coord}(\mathcal{T}_s(x^w_{j,t}), \hat{y}^T_{j,t}) &= \sum_k^{N_{j,t}} (c_{j,t,k}^T)^2 \mathcal{L}_{coord}(\mathcal{T}_s(x^w_{j,t}), \hat{y}^T_{j,t})_k \\
\mathcal{L}_{conf}(\mathcal{T}_s(x^w_{j,t}), \hat{y}^T_{j,t}) &= \sum_k^{N_{j,t}} (c_{j,t,k}^T)^2 \mathcal{L}_{conf}(\mathcal{T}_s(x^w_{j,t}), \hat{y}^T_{j,t})_k 
\end{aligned}
\vspace{-5pt}
\end{equation}
where $N_{j,t}$ is the number of pseudo-labels in frame $t$ of video $j$. $c^T_{j,t,k}$ denotes the confidence of the $k$-th pseudo-label in a given frame. $\mathcal{L}_{coord}(\mathcal{T}_s(x^w_{j,t}), \hat{y}^w_{j,t})_k$ and $\mathcal{L}_{conf}(\mathcal{T}_s(x^w_{j,t}), \hat{y}^w_{j,t})_k $ are the bounding box coordinate and confidence loss components for the $k$-th pseudo-label.

\subsection{Bidirectional, Adaptive, Teacher-Student Mutual Refinement}
The learning dynamics between the teacher and student also play a significant role in determining training stability and model robustness. 
First, the teacher should be updated at a sufficient rate such that it can catch up to the student before the student overfits, i.e., the EMA keep rate $\alpha_e$ cannot be too large. At the same time, the teacher should be characterized by gradual changes in the training curve as opposed to rapid oscillations. Thus the teacher also cannot be made to update too quickly, i.e., $\alpha_e$ cannot be too small. Finally, the student's training curve should not have sudden drops, for example, due to a bad training batch. Here, we introduce two additional techniques to dynamically balance the rate and direction of knowledge transfer during mutual learning:


\noindent\textbf{Adaptive EMA (Student $\rightarrow$ Teacher):} When using a fixed EMA keep rate $\alpha_e$, there is a trade-off between training stability and rate of knowledge transfer from the student. Instead, we propose an adaptive EMA keep rate that is conditioned on the relative performance of the teacher and student after each training epoch. We use a sigmoid-shaped function for $\alpha_e$, given by:
\vspace{-5pt}
\begin{equation}
\alpha_e = \alpha_{e, min} + (\alpha_{e, max} - \alpha_{e, min}) \cdot \frac{1}{1+e^{-\tau_0 (m_T - m_S) - \tau_1}}
\vspace{-5pt}
\end{equation}
where $\alpha_{e, min}$, $\alpha_{e, max}$, $\tau_0$ and $\tau_1$ are hyper-parameters defining the function shape. $m_T$ and $m_S$ denote teacher and student performance on a validation set according to some evaluation metric. The adaptive scheme allows $\alpha_e$ to be dynamically adjusted, that is, $\alpha_e$ is decreased (higher rate of knowledge transfer) as the student outperforms the teacher, and increased (lower rate of knowledge transfer) as the student underperforms compared to the teacher. 

\noindent\textbf{Inverse Adaptive EMA (Teacher $\rightarrow$ Student):} To avoid sudden drops in performance by the student during mutual learning, we further introduce a mechanism which allows knowledge transfer in the reversed direction, i.e., from the teacher to the student. We design a similar sigmoidal function for the inverse EMA keep rate $\alpha_{inv}$, given by:
\vspace{-5pt}
\begin{equation}
\alpha_{inv} = 
\begin{cases}
    1, & m_T \leq m_S\\
    \alpha_{inv, min} + (2 - 2\alpha_{inv, min}) \cdot \frac{1}{1+e^{-\tau_2 (m_S - m_T)}}, & m_T > m_S
\end{cases}
\vspace{-5pt}
\end{equation}
where $\alpha_{inv, min}$ and $\tau_2$ are hyper-parameters of the function. Here, knowledge transfer to the student is increased (lower $\alpha_{inv}$) when the teacher outperforms the student, and decreased (higher $\alpha_{inv}$) when the teacher underperforms.

\section{Experiments}
\subsection{Experimental Settings}
\noindent\textbf{Data.}
An extensive retrospective, multi-center clinical dataset of 7,998 lung ultrasound videos were used in this work. The data were acquired from 420 patients with suspicion of lung consolidation or other related pathology (e.g., pneumonia, pleural effusion) from 8 U.S. clinical sites between 2017 and 2020. The videos were each at least 3 seconds in length and contained at least 60 frames. 385 (fully labeled training set), 337 (validation set) and 599 (test set) videos were annotated for lung consolidation regions using bounding boxes. All data were partitioned at the subject level. The remaining 6,677 videos were annotated only for the presence or absence of lung consolidation at the video level (weakly labeled set). Annotation was carried out by a multi-center team of expert physicians with medical training in lung ultrasound. Each video was annotated by two experts and adjudicated by a third expert when a disagreement between the first two annotators occurred. 

\noindent\textbf{Implementation Details.}
We used the PyTorch Ultralytics implementation of the YOLO-v5 object detector \cite{yolov5} with default training settings (Adam optimizer with learning rate of 0.001). The weights for the confidence and coordinate losses were set to $\lambda_{conf}=1.0$, $\lambda_{coord}=0.05$. The weights for the frame-level fully supervised, frame-level semi-supervised, and video-level weakly supervised losses were set to $\lambda_{f\_sup}=\lambda_{f\_semi}=1$, and $\lambda_{v\_weak}=0.05$. For training from pseudo-labels without re-weighting, the confidence threshold was set to $\beta=0.5$. Otherwise, hyperparameters were set to $\beta=\beta_l=0.1$ when using weighted pseudo-labels. To train with a fixed EMA keep rate, we used $\alpha_{e}=0.95$. Otherwise, when applying bidirectional adaptive EMA for mutual teacher-student refinement, the hyperparameters were set to $\alpha_{e,min}=0.75$, $\alpha_{e,max}=0.99$, $\alpha_{inv,min}=0.85$, $\tau_0=180$, $\tau_1=3$, and $\tau_2=180$.

\begin{table}[t]
\caption{Validation and test mAP for fully, semi-, and weakly semi-supervised models. Mean $\pm$ standard deviation based on five repeated experiments. All methods were significantly superior to YOLO (first row) and significantly inferior to the proposed method (last row)(paired two-way t-test, p-value $<$ 0.05)}
\vspace{-5pt}
\begin{tabular}{p{2.4cm}p{5.2cm}|p{2.4cm}p{2.2cm}}
\hline
Category & Method & Validation mAP & Test mAP\\
\hline
\multirow{2}{*}{Fully supervised}  & YOLO \cite{yolov5} & 0.435 $\pm$ 0.012 & 0.412 $\pm$ 0.013 \\ 
  & YOLO+HE & 0.452 $\pm$ 0.015 & 0.440 $\pm$ 0.017 \\ 

\hline
Semi-supervised & YOLO+HE+Unlabeled & 0.468 $\pm$ 0.005 & 0.447 $\pm$ 0.003 \\

\hline
\multirow{4}{*}{\shortstack{Weakly \\Semi-supervised}} & YOLO+HE+Weak & 0.505 $\pm$ 0.004 & 0.476 $\pm$ 0.002  \\
& YOLO+HE+Weak+Pseudo & 0.508 $\pm$ 0.004  & 0.479 $\pm$ 0.004\\
& YOLO+HE+Weak+TSMR & 0.515 $\pm$ 0.005 & 0.480 $\pm$ 0.004\\
& YOLO+HE+Weak+Pseudo+TSMR & \textbf{0.519 $\pm$ 0.003} & \textbf{0.484 $\pm$ 0.003}\\
\hline
\end{tabular}
\vspace{1pt}

\vspace{-10pt}
\label{tab1}
\end{table}

\noindent\textbf{Experiments.}
We first trained a baseline YOLO-v5 detector on the fully supervised data $\mathcal{D}_f$, denoted as \textbf{YOLO}. We compared the baseline to a fully supervised training experiment with hierarchical (iteration and epoch-based) EMA training during burn-in, as proposed in Section 3.2; this is denoted by \textbf{+HE}. All subsequent experiments involving teacher-student mutual learning were initialized from the same YOLO+HE model checkpoint. We implemented the semi-supervised approach from Unbiased Teacher \cite{liu2021unbiased} by adding all videos from the weakly supervised dataset $\mathcal{D}_w$, but without providing video-level labels (i.e., treating these as unlabeled data); this is denoted as \textbf{+Unlabeled}. Note, to demonstrate the effectiveness of the proposed method, +Unlabeled a derived version of the Unbiased Teacher \cite{liu2021unbiased}, using its way of utilizing unlabeled data while keeping the same setting of the model and train strategy as the rest of competing methods. We then introduced our proposed method of weak semi-supervision, described in Section 3.3, by including the video-level labels from $\mathcal{D}_w$, denoted as \textbf{+Weak}. Finally, experiments using our methods for pseudo-label re-weighting (Section 3.4) and bidirectional, adaptive, teacher-student mutual refinement (Section 3.5) are denoted as \textbf{+Pseudo} and \textbf{+TSMR} respectively. We compared mean Average Precision (mAP) on the validation and test sets described above, with each experiment repeated five times to assess repeatability. Experimental results are summarized in Table \ref{tab1}.

\subsection{Results \& Discussion}


\begin{figure}[t]
\centering
\includegraphics[width=0.9\textwidth]{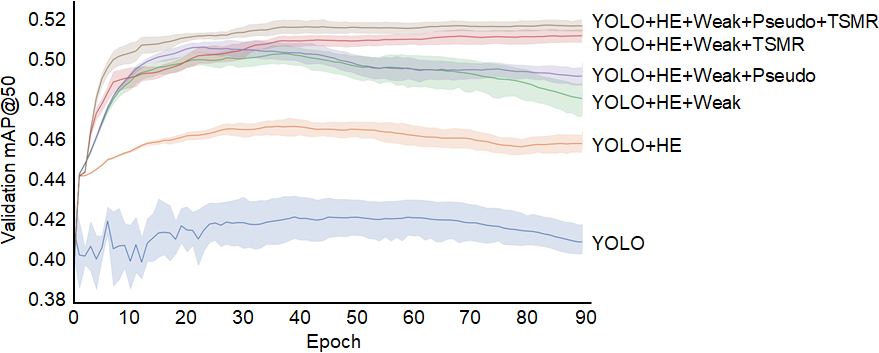}
\vspace{-10pt}
\caption{Learning curves of teacher models $\theta_T$ on validation set during teacher-student mutual learning. Solid lines show mean validation mAP across five repeated experiments. Ranges indicate 95\% confidence intervals.} 
\label{fig:5_repeat}
\vspace{-5pt}
\end{figure}

\noindent\textbf{Contribution of Hierarchical EMA Training:}
The baseline, fully-supervised detector (YOLO) achieved validation and test mAP of 0.435 and 0.412, respectively. These improved to 0.452 and 0.440, respectively, with the inclusion of the hierarchical EMA training strategy during burn-in (YOLO+HE). 

\begin{figure}[t]
\centering
\includegraphics[width=0.6\textwidth]{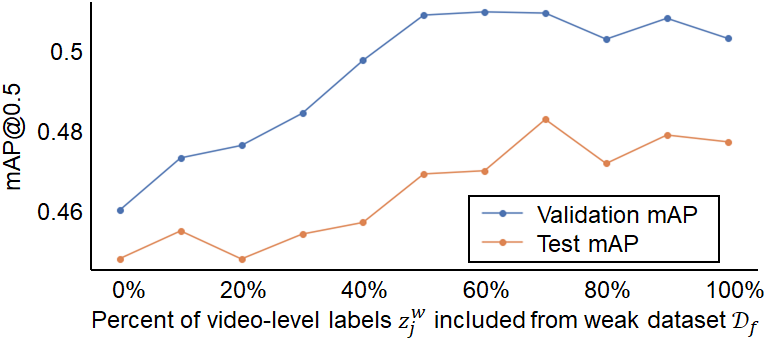}
\vspace{-10pt}
\caption{Contribution of video-level supervision during teacher-student mutual learning. Models were trained with varying proportions of video labels relative to unlabeled data.} 
\label{fig:fractional}
\vspace{-5pt}
\end{figure}

\noindent\textbf{Contribution of Semi- and Weak (Video-level) Supervision:}
The addition of unlabeled data and semi-supervision based on YOLO+HE+Unlabeled improved validation mAP from 0.452 to 0.468 and test mAP from 0.440 to 0.447, which was a statistically significant increase (p-value $<$ 0.05). Furthermore, the standard deviation of mAP values over repeated experiments decreased (0.015 to 0.005 for validation, 0.017 to 0.003 for test), suggesting that the semi-supervised model is more stable and repeatable across runs. This was also reflected in the tighter 95\% confidence intervals for validation mAP learning curves across repeated runs (Fig. \ref{fig:5_repeat}). Model performance again increased with the introduction of weak supervision of video-level labels (YOLO+HE+Weak) (validation mAP 0.505, test mAP 0.476, p-value $<$ 0.05), with corresponding decreases in mAP standard deviation (0.004 in validation, 0.002 in test) and 95\% confidence intervals over repeat runs (Fig. \ref{fig:5_repeat}). To further investigate the contribution of video-level supervision, we trained models with all fully labeled data $\mathcal{D}_f$ but utilized a proportion of video labels $z^w_j$ with the remainder of $\mathcal{D}_w$ treated as unlabeled. mAP improved consistently with increased video-level supervision (Fig. \ref{fig:fractional}). 

\noindent\textbf{Contribution of Weighted Pseudo-labels:}
The use of weighted pseudo-labels (YOLO+HE+Weak+Pseudo) further improved validation mAP from 0.505 to 0.508 and test mAP from 0.476 to 0.479. The re-weighting mechanism eliminated all false positive pseudo-labels in negative videos and increased the number of pseudo-labels to better match the overall number of true labels. In comparison, fixed pseudo-label thresholds resulted in worse detection performance (test mAP 0.477, 0.474, 0.476, and 0.474 for thresholds of 0.1, 0.3, 0.5, and 0.7). 


\noindent\textbf{Contribution of Bidirectional Teacher-Student Mutual Refinement:}
Teacher-student mutual refinement (YOLO+HE+Weak+TSMR) further boosted validation mAP from 0.508 to 0.515 and test mAP from 0.479 to 0.480. Model repeatability was improved, as shown in Fig. \ref{fig:5_repeat}, where variation between experiments was greatly reduced (narrow 95\% confidence intervals) and model convergence occurred more quickly. Ablation experiments confirmed that a fixed EMA keep rate $\alpha_e$ was unable to achieve comparable detection performance compared to the proposed bidirectional adaptive EMA updates ($\alpha_e=0.9$, lower fixed rate: 0.506 and 0.471; $\alpha_e=0.95$, moderate fixed rate: 0.505 and 0.476; and $\alpha_e=0.99$, high fixed rate: 0.489 and 0.470, for validation mAP and test mAP respectively).

\begin{figure}[t]
\centering
\includegraphics[width=0.8\textwidth]{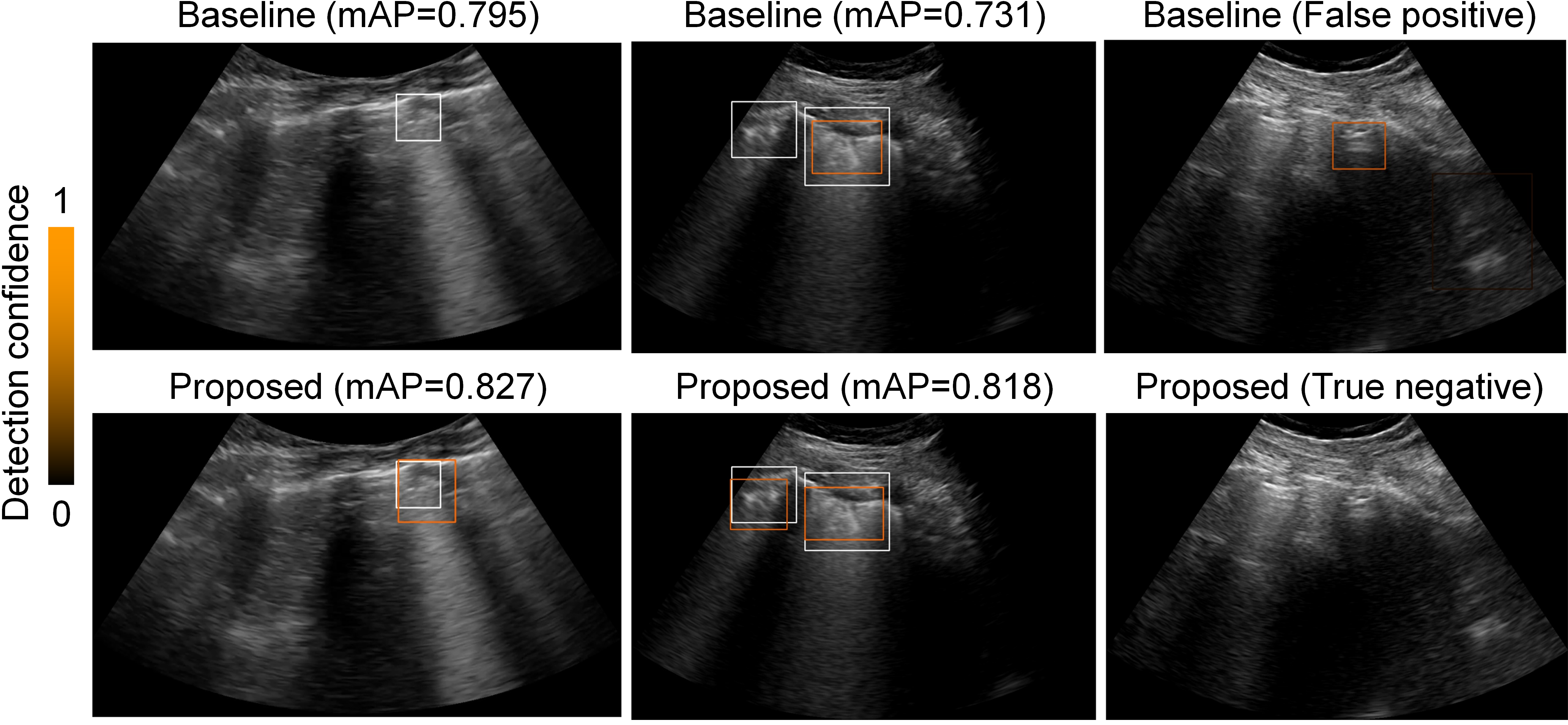}
\vspace{-10pt}
\caption{Lung consolidation detection with baseline YOLO (top) and proposed method (YOLO+HE+Weak+Pseudo+TSMR, bottom). White boxes show expert annotations. Orange boxes show model detections. mAPs calculated for each video are also shown.} 
\label{fig:qual_example}
\vspace{-5pt}
\end{figure}

\noindent\textbf{Final Results for Proposed Method:}
Finally, best-performing models incorporating all proposed components achieved validation mAP of 0.519 and test mAP of 0.484, which were statistically significant improvements to both fully supervised (0.452 and 0.440) and semi-supervised (0.468 and 0.447) baselines (p-value $<$ 0.05). Furthermore, an extra experiment suggested that comparable test mAP of baseline YOLO detector (0.412) could be achieved by the proposed method using merely one-third of the labeled data (0.395). The reference speed was 2.9ms per frame on a NVIDIA GeForce RTX 3090 GPU, enabling real-time detection. 
Examples of lung consolidation detection in ultrasound are seen in Fig. \ref{fig:qual_example}, where the proposed method demonstrates successful detection of challenging pathology not identified (or falsely identified) by the baseline YOLO detector.  

\section{Conclusion}
This is the first study to introduce a weakly semi-supervised framework for object detection on medical video data. Our method extends a teacher-student training strategy to provide weak supervision via a video-level loss. We also introduce improvements to the underlying teacher-student mutual learning mechanism, including methods to improve the quality of pseudo-labels and optimize knowledge transfer between the student and teacher. Empirical results on a lung ultrasound pathology detection task demonstrate that the framework leads to improved detection accuracy and robustness compared to existing baseline models, while also being more efficient in data and annotation usage. One limitation of the method is the need to empirically select hyperparameters, which could be resolved using adaptive hyperparameter tuning techniques as part of future work. Moreover, we considered the proposed components as orthogonal research directions as other pseudo-label refinement techniques in the state-of-the-art semi-supervised detection methods, which could be further combined in achieving better performance. Lastly, the proposed improvements to the teacher-student mechanism could potentially be adapted for other semi-supervised and weakly supervised learning tasks, including classification and segmentation.

\vspace{-10pt}

\subsubsection{Acknowledgements} 
We would like to acknowledge the contributions from the following people for their efforts in data curation and annotations: Zohreh Laverriere, Xinliang Zheng (Lia), Annie Cao, Katelyn Hostetler, Yuan Zhang, Amber Halse, James Jones, Jack Lazar, Devjani Das, Tom Kennedy, Lorraine Ng, Penelope Lema, Nick Avitabile.
%
%
%
\bibliographystyle{splncs04}
\bibliography{mybib}

\begin{thebibliography}{10}
\providecommand{\url}[1]{\texttt{#1}}
\providecommand{\urlprefix}{URL }
\providecommand{\doi}[1]{https://doi.org/#1}

\bibitem{bakalo2021weakly}
Bakalo, R., Goldberger, J., Ben-Ari, R.: Weakly and semi supervised detection
  in medical imaging via deep dual branch net. Neurocomputing  \textbf{421},
  15--25 (2021)

\bibitem{dlus3}
Bassiouny, R., Mohamed, A., Umapathy, K., Khan, N.: {An Interpretable Object
  Detection-Based Model For The Diagnosis Of Neonatal Lung Diseases Using
  Ultrasound Images}. In: 2021 43rd Annual International Conference of the IEEE
  Engineering in Medicine {\&} Biology Society (EMBC). pp. 3029--3034. IEEE
  (nov 2021). \doi{10.1109/EMBC46164.2021.9630169},
  \url{https://ieeexplore.ieee.org/document/9630169/}

\bibitem{chai2022orf}
Chai, Z., Lin, H., Luo, L., Heng, P.A., Chen, H.: Orf-net: Deep omni-supervised
  rib fracture detection from chest ct. In: International Conference on Medical
  Image Computing and Computer-Assisted Intervention. pp. 238--248. Springer
  (2022)

\bibitem{ge2022point}
Ge, Y., Zhou, Q., Wang, X., Shen, C., Wang, Z., Li, H.: Point-teaching: Weakly
  semi-supervised object detection with point annotations. arXiv preprint
  arXiv:2206.00274  (2022)

\bibitem{yolov5}
Glenn~Jocher, e.a.: {ultralytics/yolov5: v6.0 - YOLOv5n 'Nano' models, Roboflow
  integration, TensorFlow export, OpenCV DNN support} (Oct 2021).
  \doi{10.5281/zenodo.5563715}

\bibitem{jeong2019consistency}
Jeong, J., Lee, S., Kim, J., Kwak, N.: Consistency-based semi-supervised
  learning for object detection. Advances in neural information processing
  systems  \textbf{32} (2019)

\bibitem{jeong2021interpolation}
Jeong, J., Verma, V., Hyun, M., Kannala, J., Kwak, N.: Interpolation-based
  semi-supervised learning for object detection. In: Proceedings of the
  IEEE/CVF Conference on Computer Vision and Pattern Recognition. pp.
  11602--11611 (2021)

\bibitem{ji2022point}
Ji, H., Liu, H., Li, Y., Xie, J., He, N., Huang, Y., Wei, D., Chen, X., Shen,
  L., Zheng, Y.: Point beyond class: A benchmark for weakly semi-supervised
  abnormality localization in chest x-rays. In: International Conference on
  Medical Image Computing and Computer-Assisted Intervention. pp. 249--260.
  Springer (2022)

\bibitem{Jiao2022}
Jiao, R., Zhang, Y., Ding, L., Cai, R., Zhang, J.: {Learning with Limited
  Annotations : A Survey on Deep Semi-Supervised Learning for Medical Image
  Segmentation}. arXiv pp. 1--19 (2022)

\bibitem{dlus1}
Kulhare, S., Zheng, X., Mehanian, C., Gregory, C., Zhu, M., Gregory, K., Xie,
  H., McAndrew~Jones, J., Wilson, B.: Ultrasound-based detection of lung
  abnormalities using single shot detection convolutional neural networks. In:
  Simulation, image processing, and ultrasound systems for assisted diagnosis
  and navigation, pp. 65--73. Springer (2018)

\bibitem{liu2021unbiased}
Liu, Y.C., Ma, C.Y., He, Z., Kuo, C.W., Chen, K., Zhang, P., Wu, B., Kira, Z.,
  Vajda, P.: Unbiased teacher for semi-supervised object detection. arXiv
  preprint arXiv:2102.09480  (2021)

\bibitem{meethal2022semi}
Meethal, A., Pedersoli, M., Zhu, Z., Romero, F.P., Granger, E.: Semi-weakly
  supervised object detection by sampling pseudo ground-truth boxes. arXiv
  preprint arXiv:2204.00147  (2022)

\bibitem{Peng2021}
Peng, J., Wang, Y.: {Medical Image Segmentation with Limited Supervision : A
  Review of Deep Network Models}. arXiv pp. 1--24 (2021)

\bibitem{ren2020ufo}
Ren, Z., Yu, Z., Yang, X., Liu, M.Y., Schwing, A.G., Kautz, J.: Ufo2: A unified
  framework towards omni-supervised object detection. In: European Conference
  on Computer Vision. pp. 288--313. Springer (2020)

\bibitem{shao2022deep}
Shao, F., Chen, L., Shao, J., Ji, W., Xiao, S., Ye, L., Zhuang, Y., Xiao, J.:
  Deep learning for weakly-supervised object detection and localization: A
  survey. Neurocomputing  (2022)

\bibitem{tang2021humble}
Tang, Y., Chen, W., Luo, Y., Zhang, Y.: Humble teachers teach better students
  for semi-supervised object detection. In: Proceedings of the IEEE/CVF
  Conference on Computer Vision and Pattern Recognition. pp. 3132--3141 (2021)

\bibitem{wang2022omni}
Wang, P., Cai, Z., Yang, H., Swaminathan, G., Vasconcelos, N., Schiele, B.,
  Soatto, S.: Omni-detr: Omni-supervised object detection with transformers.
  In: Proceedings of the IEEE/CVF Conference on Computer Vision and Pattern
  Recognition. pp. 9367--9376 (2022)

\bibitem{wang2021data}
Wang, Z., Li, Y., Guo, Y., Fang, L., Wang, S.: Data-uncertainty guided
  multi-phase learning for semi-supervised object detection. In: Proceedings of
  the IEEE/CVF Conference on Computer Vision and Pattern Recognition. pp.
  4568--4577 (2021)

\bibitem{dlus2}
Xing, W., Li, G., He, C., Huang, Q., Cui, X., Li, Q., Li, W., Chen, J., Ta, D.:
  Automatic detection of a-line in lung ultrasound images using deep learning
  and image processing. Medical Physics  (2022)

\bibitem{xu2021end}
Xu, M., Zhang, Z., Hu, H., Wang, J., Wang, L., Wei, F., Bai, X., Liu, Z.:
  End-to-end semi-supervised object detection with soft teacher. In:
  Proceedings of the IEEE/CVF International Conference on Computer Vision. pp.
  3060--3069 (2021)

\bibitem{zhang2021weakly}
Zhang, D., Zeng, W., Guo, G., Fang, C., Cheng, L., Han, J.: Weakly supervised
  semantic segmentation via alternative self-dual teaching. arXiv preprint
  arXiv:2112.09459  (2021)

\bibitem{zhang2022group}
Zhang, S., Yu, Z., Liu, L., Wang, X., Zhou, A., Chen, K.: Group r-cnn for
  weakly semi-supervised object detection with points. In: Proceedings of the
  IEEE/CVF Conference on Computer Vision and Pattern Recognition. pp.
  9417--9426 (2022)

\bibitem{zhou2021instant}
Zhou, Q., Yu, C., Wang, Z., Qian, Q., Li, H.: Instant-teaching: An end-to-end
  semi-supervised object detection framework. In: Proceedings of the IEEE/CVF
  Conference on Computer Vision and Pattern Recognition. pp. 4081--4090 (2021)

\end{thebibliography}

\end{document}